\documentclass[aps,prx,twocolumn,showpacs,superscriptaddress]{revtex4-1}

\usepackage{amssymb}
\usepackage{amsfonts}
\usepackage{amsmath}
\usepackage{bm}
\usepackage{textcomp}
\usepackage{color}
\usepackage{longtable}
\usepackage{graphicx}% Include figure files
\usepackage{dcolumn}% Align table columns on decimal point
\usepackage[a4paper=true,pagebackref=false]{hyperref}
\usepackage[normalem]{ulem}

\begin{document}

\title{Stochastic thermal feedback in switching measurements of\\superconducting nanobridge caused by overheated electrons and phonons}

\author{M.~Zgirski}
\email{zgirski@ifpan.edu.pl}
\affiliation{Institute of Physics, Polish Academy of Sciences, Aleja Lotnikow 32/46, PL 02668 Warsaw, Poland}

\author{M.~Foltyn}
\affiliation{Institute of Physics, Polish Academy of Sciences, Aleja Lotnikow 32/46, PL 02668 Warsaw, Poland}

\author{A.~Savin} \affiliation{Low Temperature Laboratory, Department of Applied Physics, Aalto University School of Science, FI-00076 Aalto, Finland}\affiliation{QTF Centre of Excellence, Department of Applied Physics, Aalto University School of Science, FI-00076 Aalto, Finland}

\author{K.~Norowski}
\affiliation{Institute of Physics, Polish Academy of Sciences, Aleja Lotnikow 32/46, PL 02668 Warsaw, Poland}

\date{\today}

\begin{abstract}
We study correlated switchings of superconducting nanobridge probed with train of current pulses. For pulses with low repetition rate each pulse transits the superconducting bridge to normal state with probability $P$ independent of the outcomes in the preceding pulses. We show that with reduction of the time interval between pulses long range correlation between pulses occurs: stochastic switching in a single pulse rises temperature of the bridge and affects outcome of the probing for next pulses. As a result, an artificial intricate stochastic process with adjustable strength of correlation is produced. We identify regime where apparent switching probability exhibits the thermal hysteresis with discontinuity at a critical current amplitude of the probing pulse. This engineered stochastic process can be viewed as an artificial phase transition and provides an interesting framework for studying correlated systems. The process resembles the familiar transition from superconducting to normal state in the current-bias nanowire, proceeding through phase slip avalanche. Due to its extreme sensitivity on the control parameter, i.e. electric current, temperature or magnetic field, it offers opportunity for ultra-sensitive detection.
\end{abstract}

\maketitle

\section{Introduction}
There are processes in nature whose understanding can be facilitated by constructing an artificial systems which exhibit very similar, often the same dynamics, expressed in the analogous equations. It is the case for mechanical model of the RF-SQUID operation\cite{Lounasmaa1974} or in studying quantum chaos by using microwave networks\cite{Sirko2019}. A great deal of understanding of the actual physical system can be acquired from such studies. In the present work we present construction and validation of the artificial process which mimics multi-phase slip or phase diffusion behavior responsible for the switching of superconducting wires and junctions, respectively. Alternatively, one may envisage it as a useful tool to get some insight into processes exhibiting abrupt transition between states due to unlikely strong correlated fluctuation. Such transition may, for example, lie at the heart of the biological evolution leading to appearance of new species on Earth\cite{Chance2015}.

Measurements of Josephson junction (JJ) switching are widely used to study decay of metastable states, when junction transits from superconducting to finite voltage state\cite{Fulton1974,Weiss2017,Clarke1985,Martinis1988,Hanggi1990}. They allow for experimental determination of the current-phase relation\cite{Zgirski2011} and find application in sensing tiny magnetic moments if two junction are connected to form a SQUID loop\cite{Wernsdorfer2009}. They are used for read-out of superconducting qubits\cite{Chiorescu2003}, detecting electromagnetic noise\cite{LeMasne2009} or single photons\cite{Walsh2017}. Measurement of the switching probability as a function of biasing current yields S-shaped curve. Its exact shape bears information about fluctuation mechanism underlying the escape process. For underdamped junctions the escape is triggered by a single thermal fluctuation or macroscopic quantum tunneling event that leads to running of the superconducting phase down the tilted washboard potential\cite{Clarke1985,Martinis1988}. However, for a junction in the moderately damped regime transition from superconducting to normal state cannot be explained by assumption of a single fluctuation and instead involves phase diffusion eventually leading to escape\cite{Delsing2005,Pekola2005,Kautz1990,Massarotti2015}. For wires at temperatures sufficiently close to $T_c$ similar intricate stochastic process is predicted: the switching to normal state is the result of consecutive phase slip events with each phase slip increasing the probability for the next phase slip to occur\cite{Bezryadin2009,Li2011,Murphy2015,Baumans2017}. It is the experimental emulation of this last process that we present in our work: we consider the process of the consecutive switching events in a regime where each switching event increases the probability for the next switching event to occur. We need to stress that our presentation does not investigate the switching mechanism of the bridge tested with current pulses. Whatever this mechanism is, we will use the fact that it transits the bridge to the normal state with the local electron temperature larger than $T_c$.

The independent switching experiments manifest themselves in the binomial distribution of switchings for a fixed value of the probing current: JJ behaves like a coin for which head and tail experiment is performed\cite{Zgirski2019}. By virtue of independence switching probability is not affected by the outcome of the probing in any other pulse. An intriguing situation arises if we introduce correlation between switching events by reducing the time interval between probing pulses. In such case after successful switching event the subsequent thermal relaxation is not complete when the next testing pulse arrives. Thus the switching probability is enhanced. As we report below, it leads to emergence of the correlated stochastic process with the tunable strength of correlation. We demonstrate, for the first time, hysteretic switching current dependencies exhibiting the metastable behaviour for a critical value of the probing current. We argue that this behaviour is desired for threshold detection of various physical signals and may emulate natural stochastic processes, involving the phase-slip avalanche which is possibly responsible for the superconducting-normal-state transition in nanowires\cite{Bezryadin2009}.

\begin{figure}[t]
\centering
\includegraphics[width=0.5\textwidth]{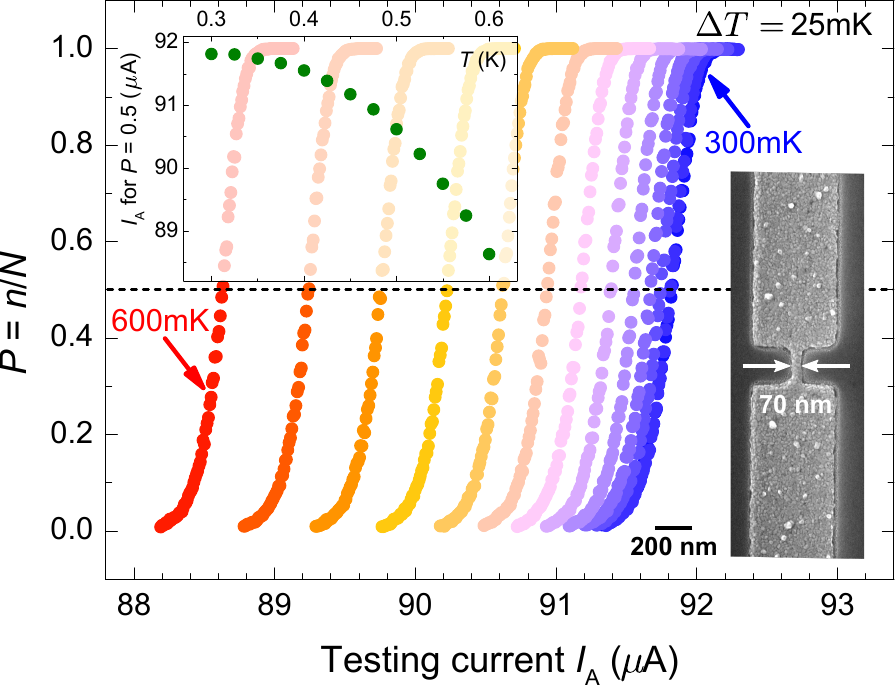}
\caption{\label{fig:Calib}JJ switching probabilities (\textsf{S} curves) measured at different bath temperatures used to extract switching current dependence on temperature presented in the inset. Dashed line corresponds to $P$ = 0.5. The SEM image of the aluminum nanobridge is shown on the right-hand side.}
\end{figure}

\begin{figure}[t]
\centering
\includegraphics[width=0.5\textwidth]{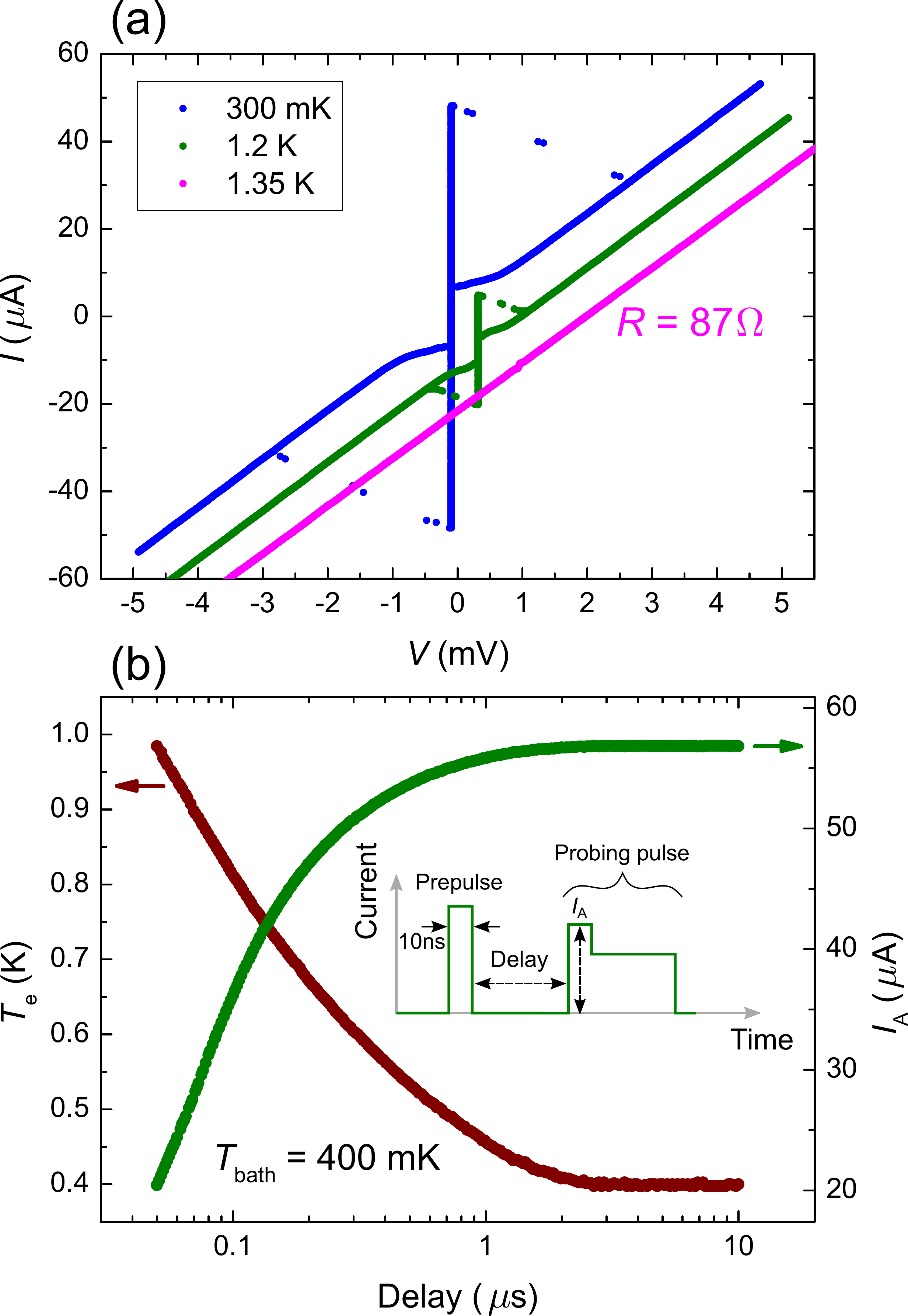}
\caption{\label{fig:Switchingcurrent}The thermal dynamics of a typical Dayem nanobridge. (a) The current-voltage characteristics of a bridge,
revealing switching behavior at a threshold current. The I-V curves are obtained at three temperatures: 300$\,$mK, 1.2$\,$K, and 1.35$\,$K (from left to right, offset horizontally for clarity). $R=87\,\Omega$, i.e. the inverse slope of all IV curves is the same as the normal state resistance of the sample (nanobridge + long nanowire) measured above $T_c=1.3\,$K. (b) Relaxation of the switching current after overheating the bridge with a heating pulse above $T_c$ (right Y axis). The corresponding temporal variation of electron temperature $T_e$ (left Y axis). The definition of the testing sequence consists of 10$\,$ns-heating prepulse followed by the probing pulse (inset). The probing pulse consists of a short part of higher amplitude intended to test the junction, and much longer sustain part enabling the read-out of the state of the bridge with low-pass-filtered line. For details see ref.$\,$\cite{Zgirski2018}, Fig.\,3.}
\end{figure}

\begin{figure*}
\centering
\includegraphics[width=0.95\textwidth]{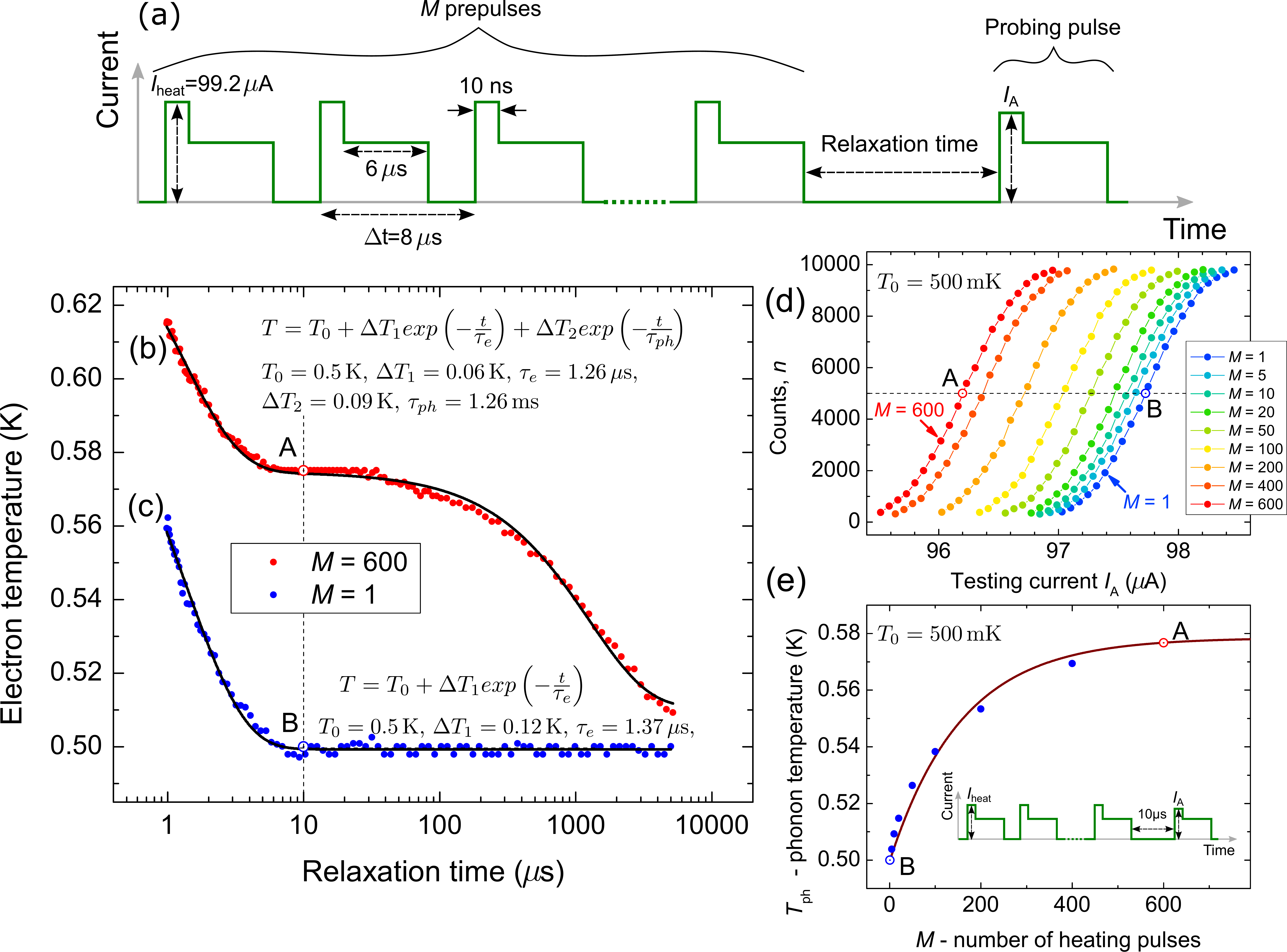}
\caption{\label{fig:Relaxation}The thermal dynamics of the nanobridge presented in the text. (a) The definition of the testing sequence. The single sequence consists of $M$ heating prepulses followed by the testing pulse. The sequence is repeated $N$ times to measure the switching number $n$. The repetition period is 10$\,$ms, to allow for proper equilibration after each sequence. (b) Relaxation of the nanobridge after $M = 600$ forced switchings: two relaxation mechanisms are visible: the fast process is the same as in (c) and the slow one is approximately 1$\,$000 times slower and is attributed to relaxation of the local phonon temperature (substrate) toward the bath temperature. The line is a fit to the sum of two exponential functions.(c) Relaxation of the excess hot-electron energy toward equilibrium with local phonons after a single forced switching ($M=1$): the local phonons are at bath temperature. The line is a fit to the single exponential function. (d) \textsf{S} curves measured for different numbers of heating prepulses 10$\,\mu$s after the end of the last prepulse, when the electrons are already thermalized at the local phonon temperature. (e) The temperature rise of the local phonons as extracted from the
\textsf{S} curves presented in (d) along the dotted line $n=5\,000$, with the aid of the calibration curve [see the inset of Fig. 1 in ref.$\,$\cite{Zgirski2019}]. The line is a single-parameter fit to Eq.\,(\ref{eq:1}) with $\Delta t=8\,\mu$s and $\tau_{ph}=1.26\,$ms yielding $\delta T_{ph}=0.50\,$mK.}
\end{figure*}

\begin{figure}
\centering
\includegraphics[width=0.48\textwidth]{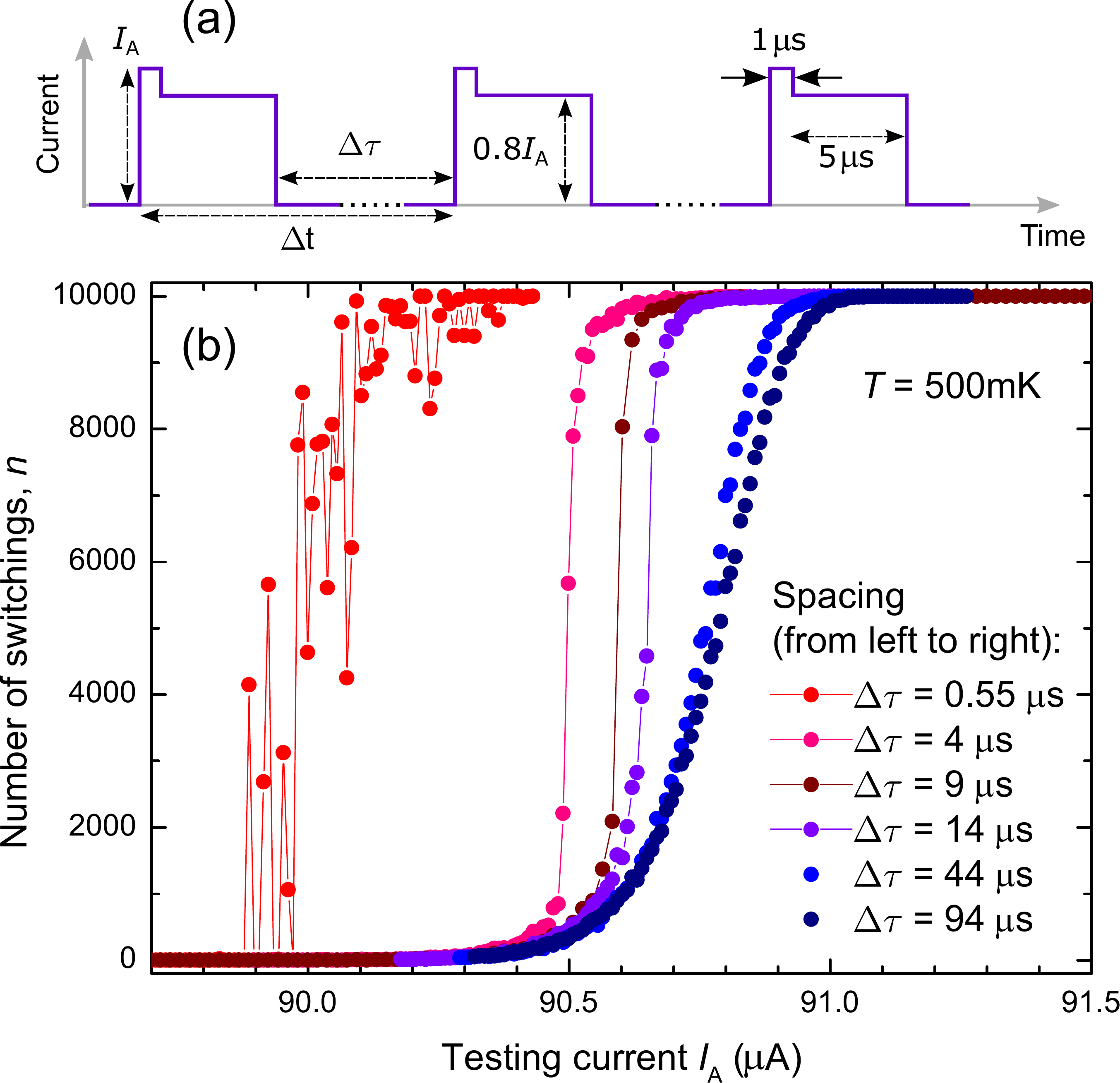}
\caption{\label{fig:Scurves}(a) Train of $N$ current pulses used to probe the bridge with different $\Delta \tau$. (b) Current dependencies of the switching number $n$ for different spacing $\Delta \tau$ between probing pulses. Each point corresponds to a finite-length train measurement ($N$=10\,000).}
\end{figure}

\section{Switching thermometry} \label{thermometry}
We fabricate a superconducting nanobridge (width = 70$\,$nm) interrupting in the middle a long nanowire (width = 600$\,$nm, length = 100$\,\mu$m) connected to large area contact pads at both sides [inset of Fig.\,\ref{fig:Calib}]. The structure is prepared by means of conventional one step e-beam lithography followed by evaporation of 30$\,$nm thick aluminum on the 300$\,\mu$m thick silicon substrate. The IV curve of the bridge reveals pronounced thermal hysteresis: the bridge switches to the normal state at a specific current value known as the switching current $I_{sw}$, and retraps back to the superconducting state only after the applied current $I_A$ is sufficiently reduced. Such behavior is explained by self-heating of the sample. During switching nanobridge forms a hot spot, from which the normal region expands towards contact pads: after a few tens of nanoseconds the bridge and nanowire are all above $T_c$. While the nanobridge resistance is only 2$\,\Omega$ the normal branch of the IV curves, measured for bath temperatures $T_0<T_c$, shows the normal state resistance of the whole sample, i.e. nanobridge with connecting nanowires [Fig.\,\ref{fig:Switchingcurrent}(a)]. Noteworthy, the bath temperature $T_0$, as read-out with a resistive RuO$_x$ thermometer, remains completely unaltered, but the local temperature of the nanobridge and nanowire rises above $T_c$. Only lowering the applied current $I_A$ leads to reduction of both the Joule heating and the related electron temperature $T_e$ and allows the sample to enter the superconducting state again when $I_A<I_{sw}(T_e)$. Our interpretation goes in line with ref.$\,$\cite{Pekola2008}, which proofs the overheating of electrons in the proximity junctions after switching. Also, for MoGe superconducting nanowires the hysteresis in IV curve is explained by self-heating\cite{Tinkham2003}, and not by the dynamics of phase motion in a tilted washboard potential, as often assumed. In our case, the switching current is 1-2 orders of magnitude higher than in the discusses references and thus, heating effects are expected to be much more pronounced. Recently, correlations in switching events for Nb-Al-Nb proximity junction have been reported, and attributed to thermal dynamics of electrons in the junction\cite{Spahr2020}.

We test the bridge with train of $N$ current pulses. In response to each pulse, dependently on the probing current amplitude and fluctuations, the bridge may remain in the superconducting state or transit to a normal state in the process known as switching. For low probing currents bridge never switches, for high current it always switches. At the intermediate values of current the bridge switches on average $n$ times. The ratio $n/N$ is an estimator for switching probability $P$ and renders familiar S-shaped curve as probing current increases [Fig.\,\ref{fig:Calib}]. The switching current dependence on temperature [see the inset in Fig.\,\ref{fig:Calib}] makes the bridge fast and sensitive thermometer\cite{Zgirski2018}, as it is evident from mutual shift of \textsf{S} curves measured at different temperatures [Fig.\,\ref{fig:Calib}].

To measure thermal relaxation of the bridge we send a heating prepulse with the amplitude exceeding the switching threshold. This forced switching transits the bridge to normal state with electron temperature larger than $T_c$. The subsequent relaxation is probed with the delayed probing pulse. The sequence is repeated $N=10\,000$ times with period long enough to provide thermalization of the bridge at the bath temperature $T_0$ after each sequence. The current amplitude of the probing pulse is adjusted to yield $P=0.5$ for each delay. The typical relaxation profile of the switching current and the resulting temperature evolution are presented in Fig.\,\ref{fig:Switchingcurrent}(b). For the more detailed description of the method please refer to our earlier works\cite{Zgirski2015, Zgirski2018, Zgirski2019, Zgirski2020}.

\section{Two thermal relaxation times} \label{tworelaxations}
We use the switching thermometry to analyze the influence of $M$ heating prepulses on the temperature of the bridge and define dynamic thermal processes involved in the relaxation of electron temperature towards bath temperature $T_0$. The experimental data presented in this paragraph are widely described in Sec.\,V of the ref.$\,$\cite{Zgirski2019}. For completeness of our presentation we analyze this data once again in the current manuscript introducing as a new element quantification of the phonon overheating due to single switching event.

We intentionally heat the bridge with the set of $M$ prepulses preceding the actual testing pulse and vary the time between the last prepulse and the probing pulse [Fig.\,\ref{fig:Relaxation}(a)]. The current amplitude for the prepulses is set constant, at a level exceeding the switching threshold, yielding forced switching in each prepulse. After a single prepulse ($M=1$), the electron temperature of the nanobridge exceeds $T_C$ and relaxes to the local phonon temperature on a time scale of a few microseconds [Fig.\,\ref{fig:Relaxation}(c)]. This relaxation is governed by electron-phonon coupling and by hot-electron diffusion\cite{Zgirski2018,Zgirski2019}. Both energy-relaxation channels bring electrons into thermal equilibrium with phonons. As we increase the number of prepulses $M$, local phonons become overheated. This overheating relaxes at an approximately 1$\,$000 times slower rate than it takes for the hot electrons to achieve the local phonon temperature [Fig.\,\ref{fig:Relaxation}(b)]. Since the two processes exhibit such different dynamics, they can be described in the linear regime using the sum of two exponential decays, yielding the two relaxation times: $\tau_e=1.26\,\mu$s for electrons, and $\tau_{ph}=1.26\,$ms for phonons.

Such dynamic investigations of the thermal processes of nanostructures are rare in literature and limited to the lowest temperatures where they are relatively slow. Authors report direct measurements of thermal relaxation times of copper ($\tau_{Cu}=10\,\mu$s) and silver ($\tau_{Ag}=500\,$ns) at $T_0$=200$\,$mK\cite{Pekola2018}. Similar measurements for gold yields $\tau_{Au}=1.6\,\mu$s at $T_0$=300$\,$mK\cite{Cleland2004}. The relaxation time for electrons in our nanostructure, measured at $T_0$=500$\,$mK is expected to be significantly shorter than $\tau_{Al}$=30$\,\mu$s reported in ref.$\,$\cite{Klapwijk2008} for $T_0$=300$\,$mK, owing to strong suppression of electron-phonon coupling at low temperatures.

It is instructive to measure the increase in temperature of the local phonons with the number of prepulses after the fast electron-phonon relaxation is over and the slow phonon-bath relaxation has not yet started. This can be accomplished by adjusting the relaxation time to 10$\,\mu$s [see point A in Fig.\,\ref{fig:Relaxation}(b)] and the result is presented in Fig.\,\ref{fig:Relaxation}(d). Recalculation of the switching current in terms of temperature with the aid of a calibration curve [see the inset of Fig.\,1 in ref.$\,$\cite{Zgirski2019}] yields a $T_{ph}(N)$ dependence, showing a slow but monotonous increase with a tendency for saturation [Fig.\,\ref{fig:Relaxation}(e)]. Obviously, the local substrate (phonon) temperature does not relax to the bath temperature before the arrival of the next testing pulse. Each switching event deposits a bit of energy in the substrate and gives rise to $\delta T_{ph}$ increase in the phonon temperature as defined 10$\,\mu$s after switching event. The contribution from each switching event relaxes exponentially over much longer times, $\tau_{ph}=1.26\,$ms as determined from the fit in Fig.\,\ref{fig:Relaxation}(b). We describe the excess phonon temperature "seen" by ($n+1$)th probing pulse with the following recursive relation:

\begin{equation}
\label{eq:1}
\begin{aligned}
\Delta T_{ph}(n+1)=&[\Delta T_{ph}(n)+\delta T_{ph}]\,exp\left(-\frac{\Delta t}{\tau_{ph}}\right),\\
&\,\Delta T_{ph}(1)=0
\end{aligned}
\end{equation}

The formula applied to $T_{ph}(N)$ dependence of Fig.\,\ref{fig:Relaxation}(e) with $\Delta t=8\,\mu$s and $\tau_{ph}=1.26\,$ms yields $\delta T_{ph}=0.50\,$mK. 

The above model is constructed in analogy to the effusion process, in which a gas escapes from a leaking volume through a pinhole. To a good approximation, the amount of gas remaining in the container decreases with time exponentially. We assume that the temperature of the overheated phonons behaves in the same way. We thus consider the linear regime, where the departure of phonon temperature $\Delta T_{ph}$ from the bath temperature is small and the cooling power for the phonons is proportional to $\Delta T_{ph}$. However, in our "effusion process" the fixed amount of "gas"' i.e. $\delta T_{ph}$ increment in temperature, is added at well-defined discrete time intervals.

\begin{figure*}
\centering
\includegraphics[width=0.95\textwidth]{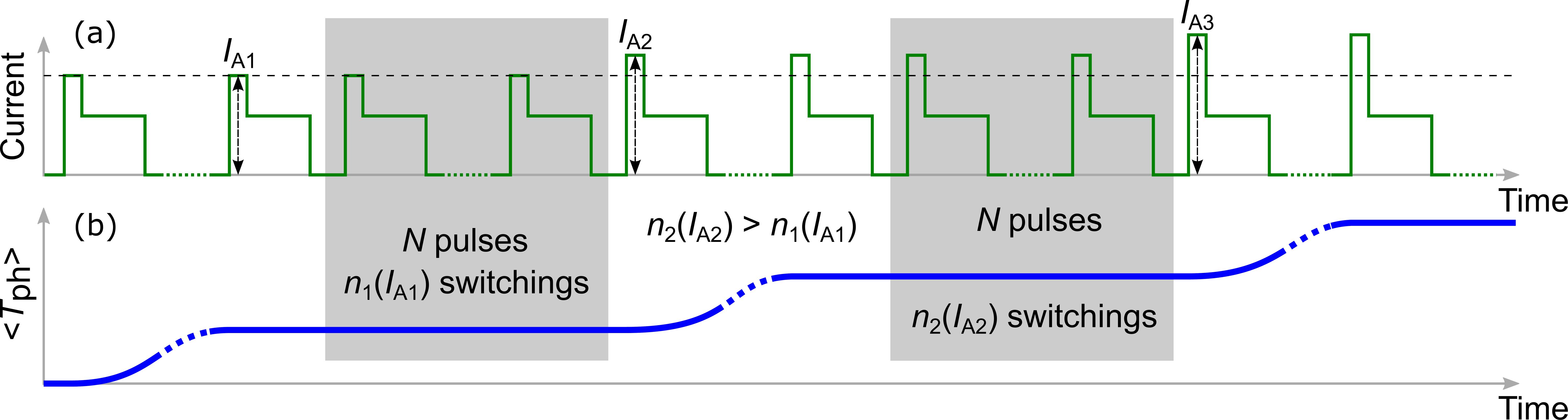}
\caption{\label{fig:Metastable_pulses} (a) The continuous train of current pulses used to probe the bridge. The testing current amplitudes $I_{A1}, I_{A2}, I_{A3}, ...$ are changed between points in step$-$like fashion preserving the continuity of the train. (b) Schematic variation of the average local phonon temperature $<T_{ph}>$ for the continuous train presented in (a). When the $I_{A}$ is increased the $<T_{ph}>$ grows slightly after every switching event to reach an elevated value after many switchings. Then the switching number is measured over temporal window (gray region) extending over $N$ pulses.}
\end{figure*}

\begin{figure*}
\centering
\includegraphics[width=0.95\textwidth]{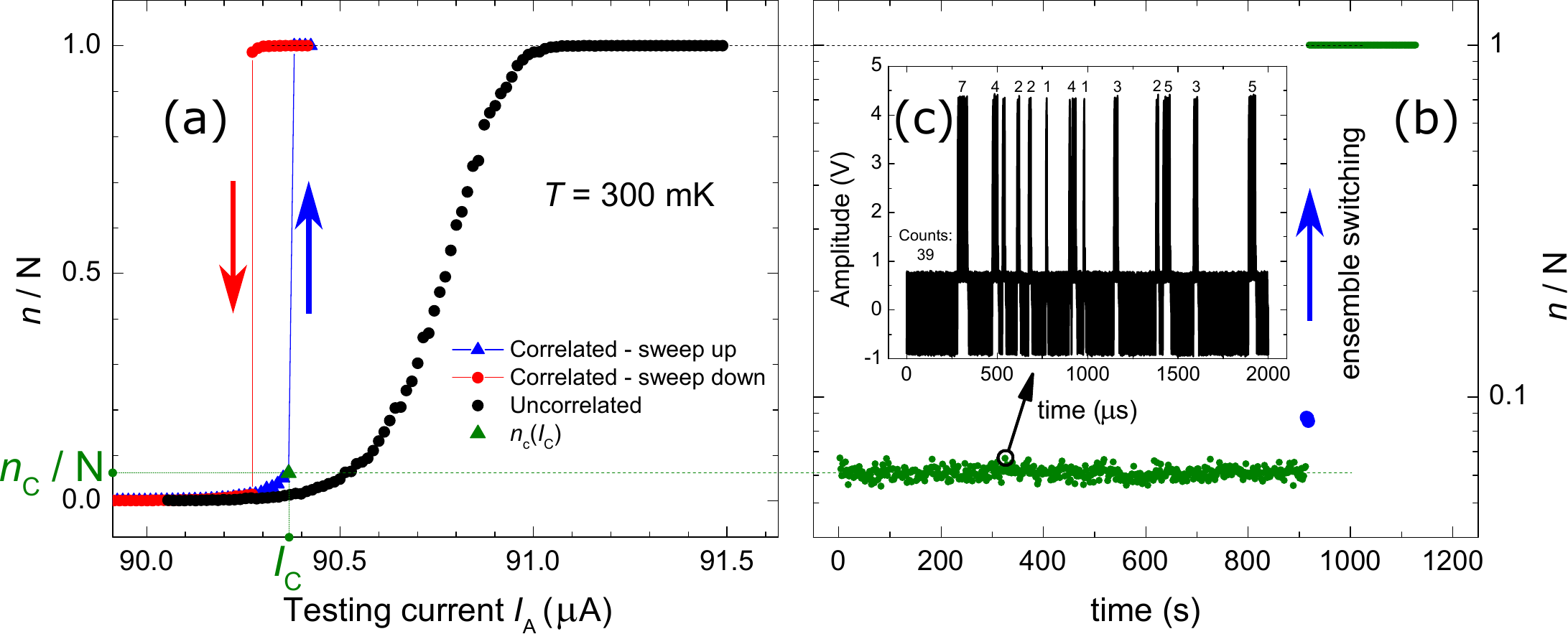}
\caption{\label{fig:Metastable}Metastable state close to transition point. (a) $n/N$ dependence on the testing current $I_A$ with abrupt transition at $I_A=I_C$ recorded for correlated pulses in the continuous train measurement. (b) The time trace of the apparent switching probability (switching number $n/N$) recorded in the neighborhood of $I_C$. The transition corresponds to the increment of the probing current by $\sim$15nA (forced transition), but it may also happen spontaneously for fixed current. Blue points correspond to a transient between two stable states, when ensemble switching happens. Each point is the result of measurement extending over time window covering $N=125\,000$ pulses with single pulse period of $\Delta t=8\mu$s and $\Delta \tau=2\mu$s. Pulse train duration is $\Delta t_{train}=N\cdot \Delta t=1sec$. (c) The discrete stochastic trajectory: an oscilloscope screenshot of 250 pulses chosen from 125$\,$000 pulses used to obtain a single point in (b) revealing bunching of switching events in a metastable state recorded while waiting for the spontaneous ensemble switching close to $I_C$. Number of switched pulses in a row is indicated above each bunch.}
\end{figure*}

\begin{figure}
\centering
\includegraphics[width=0.5\textwidth]{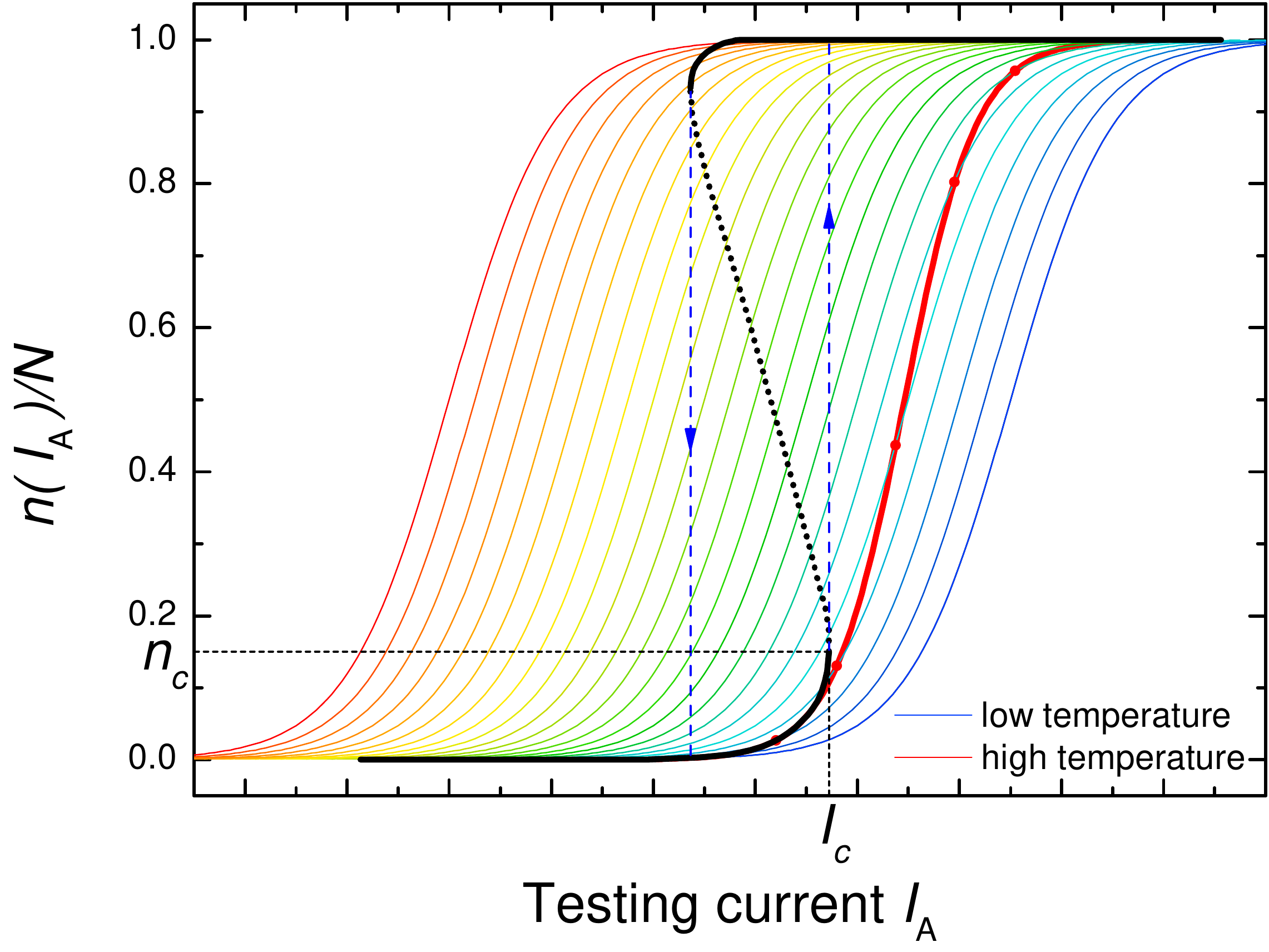}
\caption{\label{fig:backbend_scurves}Back-bent \textsf{S} curve: Schematic presentation of \textsf{S} curves for different bath temperatures in the independent switching regime [see Fig.\,\ref{fig:Calib}]. Moderate correlation leads to steepening of \textsf{S} curve like that observed in Fig.\,\ref{fig:Scurves} (red thick line). On reducing probing period discontinuous transition in the switching number $n$ is at $n=n_C$ is expected giving rise to thermally stabilized hysteresis (black curve with the dotted part not accessible to our experiment).}
\end{figure}

\begin{figure}
\centering
\includegraphics[width=0.5\textwidth]{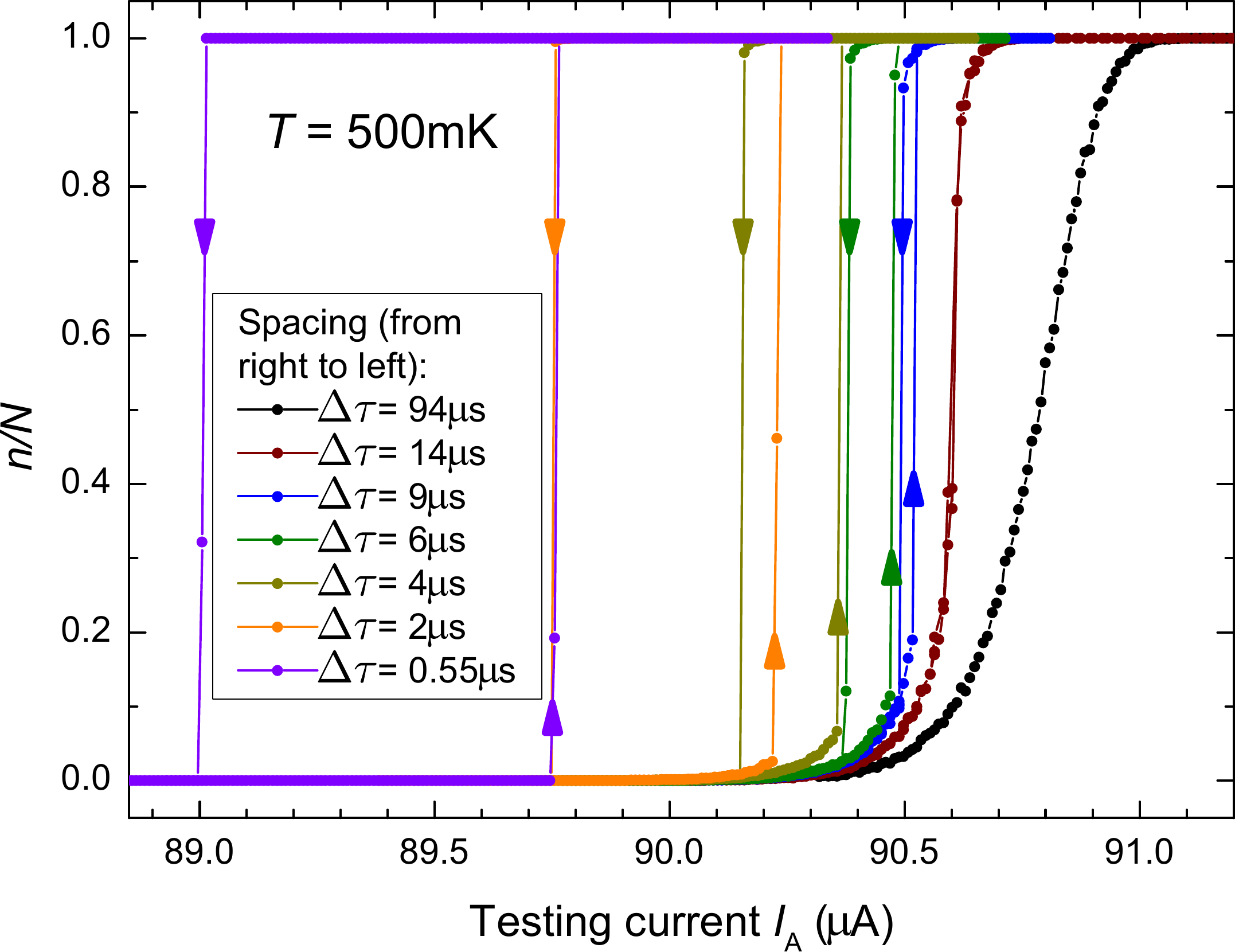}
\caption{\label{fig:Hysteresis}Experimental hysteresis loops: Probing current dependencies of the switching number $n/N$ for different temporal spacing $\Delta \tau$ between probing pulses obtained for the continuous train measurement. The rightmost curve is acquired in the independent regime. The leftmost hysteretic loop corresponds to the fully correlated regime, i.e. the panic switching\cite{Zgirski2019}.}
\end{figure}

\section{Hysteretic \textsf{S} curves and Metastable States}
In the experimental investigations presented in this paragraph we intentionally correlate switching events using time interval between pulses as a knob for controlling the strength of correlation. For pulse periods $\Delta t=100\,\mu$s no correlation is observed: switching events can be considered to be independent, for increase of $\Delta t$ above 100$\,\mu$s changes neither shape nor position of the \textsf{S} curves. With reduction of the period, switching in a single pulse start to influence the measuring result in the subsequent pulses [Fig.\,\ref{fig:Scurves}] leading to steepening of the \textsf{S} curve. The further reduction of the period destroys familiar picture of \textsf{S} curve: number of swichings $n$ corresponding to a fixed probing current amplitude seems to be completely random as it is revealed on the scattered curve presented in Fig.\,\ref{fig:Scurves} [$\Delta \tau=0.55\,\mu$s]. In fact this "noisy" data is well understood in terms of the panic switching that we studied in the ref.$\,$\cite{Zgirski2019}: the first switching appears at random, but once it happens, switchings events for all successive pulses become certain. We reserve notion of probability for independent events, while generally we will use the switching number $n/N$ specifying number of switching events in the total number $N$ of probing pulses.

As demonstrated in Sec.\,\ref{tworelaxations} sending $M$ pulses in row, each with high enough amplitude to make the bridge switch, increases the local phonon temperature. Similar effect is expected if $M$ pulses switch the bridge only with some probability. In the further studies we modify the probing protocol accordingly: the pulse train involves the thermalization pulses-intended to rise phonon temperature in the bridge to its asymptotic value for a given probing pulse amplitude, which are followed by $N=10\,000$ measuring pulses [Fig.\,\ref{fig:Metastable_pulses}].

Applying the new protocol we obtain switching dependence presented in Fig.\,\ref{fig:Metastable}. It is visible there, that switching number $n(I_A)$ exhibits abrupt transition from state $n(I_A=I_C)=n_C$ to the state $n(I_A)=N$. We call this new phenomenon the \textit{ensemble switching}, for it involves all testing pulses [Fig.\,\ref{fig:Metastable}]. Below we present qualitative explanation for the observed correlated dependency, leaving the quantitative rigorous treatment for the next paragraph. The larger $n(I_A)$ in the total number of testing pulses $N$, the higher average electron temperature in the bridge. As a result, points of higher apparent switching probability $P=n(I_A)/N$ belong to \textsf{S} curves measured at higher temperatures in the independent regime (shifted towards smaller values of the testing current) [see Fig.\,\ref{fig:Scurves}]. For moderate phonon overheating \textsf{S} curve becomes steeper, but at shorter testing periods, the increase in $n(I_A)$ above a critical value $n_C$ would be accompanied by temperature rise corresponding to smaller value of the testing current $I_A$ and back-bending of $n(I_A)$ dependence [Fig.\,\ref{fig:backbend_scurves}, black dotted curve]. Instead, since $I_A$ is increased monotonically, $n(I_A)$ exhibits abrupt transition at this point. The phenomenon has an avalanche character and relies on a thermal feedback mechanism: upon exceeding a threshold for the testing current $I_C$, small increase in the independent switching probability increases number of switchings. It, in turn, leads to increased phonon temperature thus further increasing switching probabilities. The new stable state is the one with every pulse switching the wire to the normal state. Unlike for an independent switching experiment with uncorrelated pulses, where all values of $n(I_A)$ are stable, in the correlated regime $n(I_A)$ can exhibit only values up to $n_C$ or $n=N$. Once the ensemble switching happens local phonon temperature becomes significantly elevated above bath temperature. Starting from state with $n=N$ and lowering current first we observe small suppression in the switching number $n$. It, in turn, leads to reduction of the temperature thus further reducing switching probability and transition to the state with small value of $n$. We call this transition the \textit{ensemble retrapping}.

Ramping current amplitude up and down while using the continuous pulse generation renders hysteretic loops with hysteresis more pronounced for more strongly correlated pulses [Fig.\,\ref{fig:Hysteresis}]. When the probing current amplitude is increased we see abrupt transition at $n_C$ with $n_C$ lower for shorter intervals between probing pulses. In the fully correlated regime, when the single switching triggers the ensemble switching, $n_C=0$ and $n(I_A)$ shows only two stable values: $n=0$ or $n=N$. The corresponding trace defines a rectangular hysteresis with the ensemble switching and retrapping occurring at random current values. We call such an ensemble switching in the fully correlated regime the \emph{panic switching}\cite{Zgirski2019}.

It is important to emphasis that for the hysteretic loop to be observable, the continuous pulse train must be used. If a finite-length pulse train was applied (i.e. the train with $N$ pulses of the definite $I_A$ amplitude), the bridge would relax to the bath temperature prior to arrival of the next train and "the retrapping part" of the hysteretic loop would not be accessible.

\section{Modeling of the stochastic thermal feedback} \label{fullmodel}
In this paragraph we develop a mathematical description of the switching correlation in our system, providing a proper account for abrupt transitions observed in Fig.\,\ref{fig:Metastable} and Fig.\,\ref{fig:Hysteresis}. As we saw in Sec.\,\ref{tworelaxations} there are two thermal relaxation scales for the bridge. The first one is set by the electron-phonon relaxation time. It tells how fast it takes for electrons to equilibrate with the local phonons. The second time scale is set by relaxation of the local phonon temperature with the temperature of the thermal environment i.e. bath temperature $T_0$. The slowly-relaxing local phonon overheating rises the thermal asymptote for the relaxation of the electron temperature, thus influencing switching probability over many probing pulses. The general model should assume the switching probabilities dependent on the history of the system.

We analyze pulse train with $\Delta \tau=2\,\mu$s ($\Delta t=8\,\mu$s) applied to our sample at $T_0=500\,$mK [see the corresponding experimental curve in Fig.\,\ref{fig:Hysteresis} and definition of the pulse train in Fig.\,\ref{fig:Scurves}(a)]. The relaxation time of the electron temperature is $\tau=1.3\,\mu$s, as inferred from Fig.\,\ref{fig:Relaxation}(b,c). It corresponds to overheating of electrons equal to $\Delta T_e \cong 20\,$mK, which is present at $\Delta \tau=2\,\mu$s after switching off the $M$-pulse sequence. If there is another pulse send at this moment it “sees” electrons at temperature elevated $\Delta T_e$ above the phonon temperature. But if pulse is sent $\Delta t + \Delta \tau=10\,\mu$s later, electrons are already thermalized with phonons. Noteworthy, overheating of electrons with respect to local phonons, occurring in the given switching event, affects only the next probing pulse.

The second necessary ingredient of the model takes into account slow phonon relaxation demonstrated in Fig.\,\ref{fig:Relaxation}(b). It assumes that phonon temperature remains elevated due to many switchings and each switching event gives rise to overheating of phonons, which we estimate to account for $\delta T_{ph}=0.50\,$mK increase in phonon temperature for $t=10\,\mu$s after switching event [as fitted with the model presented in the Eq.\,(\ref{eq:1})]. The contribution from each switching event relaxes exponentially over much longer times, $\tau_{ph}=1.26\,$ms as inferred from Fig.\,\ref{fig:Relaxation}(b).

\begin{figure}
\centering
\includegraphics[width=0.5\textwidth]{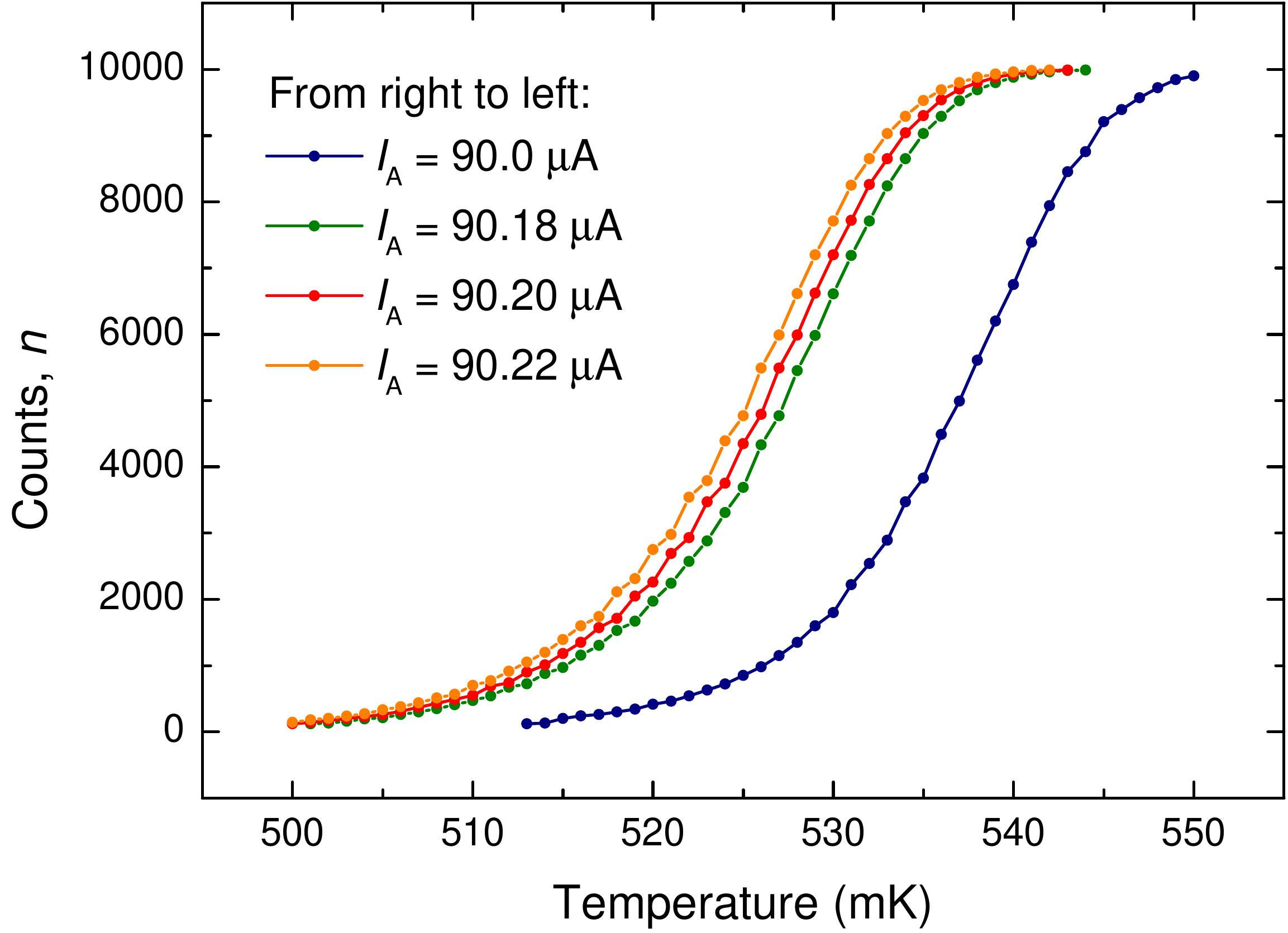}
\caption{\label{fig:Thermalscurves}Thermal \textsf{S} curves $P(T_e)=n(T_e)/N$ obtained for slightly different testing current amplitudes, as interpolated from the data presented in Fig.\,\ref{fig:Calib}.}
\end{figure}

We define the switching array $S$ consisting of ordered ones and zeros corresponding to switching event or its lack in the successive pulses, e.g. $S=[1,0,1]$ encodes switching for the first and third probing pulse and no switching in the second pulse. The corresponding switching probability $P$ in the $(n+1)$th pulse can be found from the temperature dependence of probability for a fixed probing current amplitude $P(T_e): P(n+1)=P[T_e(n+1)]$. We obtain $P(T_e)$ curves for different values of the probing current amplitude [Fig.\,\ref{fig:Thermalscurves}] by interpolating the data presented in Fig.\,\ref{fig:Calib}. We compare this probability with the randomly drawn number spanning the range from 0 to 1 to populate the $S$ array, i.e. we insert 1 into the array if the drawn number is equal or smaller than $P$, otherwise we insert 0.

\begin{figure*}
\centering
\includegraphics[width=0.95\textwidth]{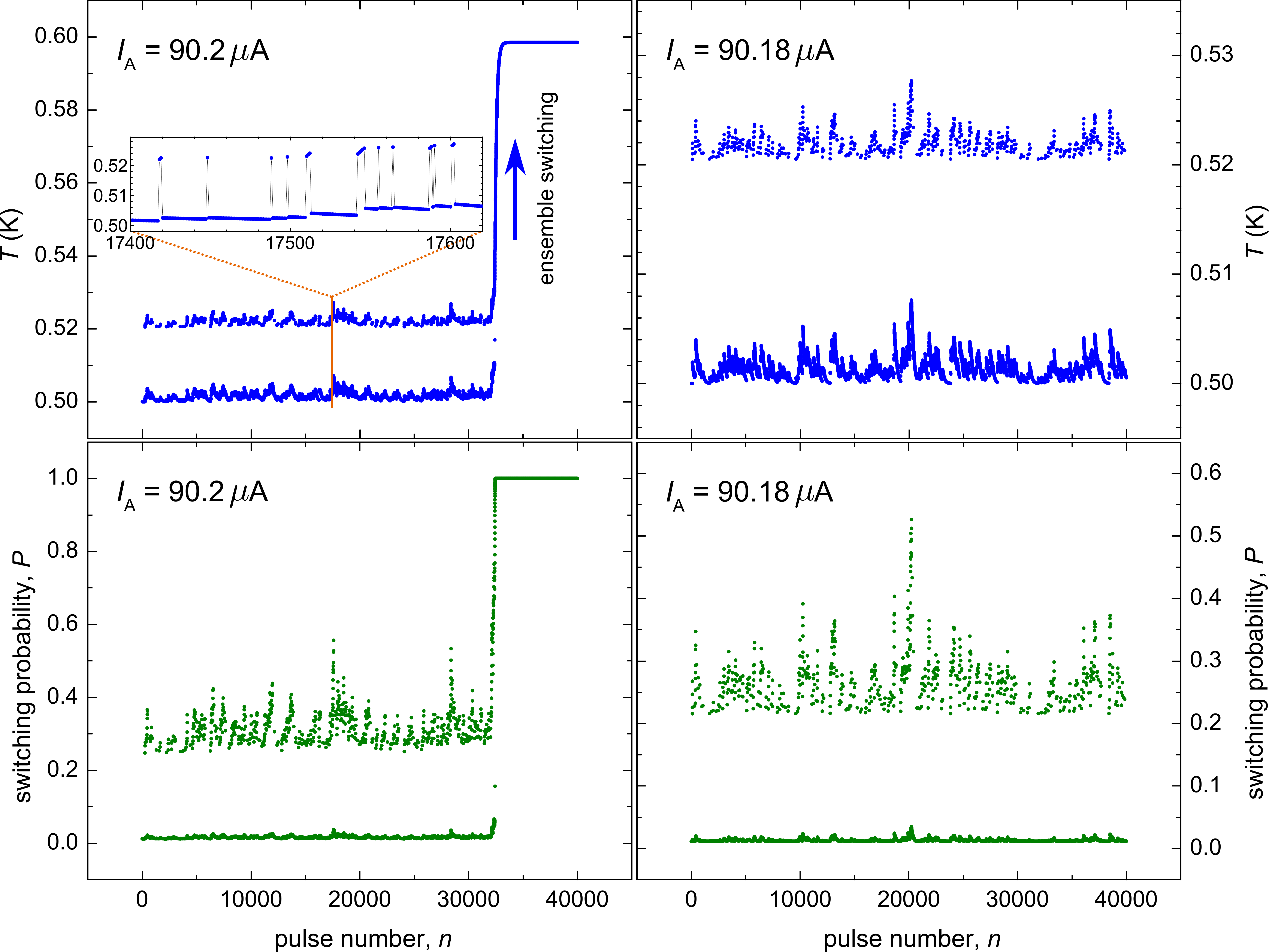}
\caption{\label{fig:Simulation}Numericaly calculated traces of electron temperature $T(n)$ [Eq.\,(\ref{eq:3})] and probability $P(n)$ obtained with aim of $P(T)$ dependence collected for $I_A=90.2\,\mu$A and 90.18$\,\mu$A. $T_0=500\,$mK. The calculation corresponds to abrupt transition observed experimentally in the hysteretic switching dependence of Fig.\,\ref{fig:Hysteresis} (curve $\Delta \tau = 2\,\mu$s). If there is a switching event for the (n-1)th pulse testing pulse, the probability or temperature for $n$th pulse adds a point to the upper branch of the corresponding trace. Otherwise, the evolution progresses through the lower branch as it is seen in the enlarged image of the temperature trace collected for $I_A=90.2\,\mu$s. Noteworthy, the trace exhibits the similar bunching of switching events prior to the ensemble switching in comparison to the experimental stochastic trajectory of Fig.\,\ref{fig:Metastable}(c). Such a dichotomy of the traces comes from the random creation of the overheated electrons in the switching event, that affects only the nearest pulse (see discussion in the text).}
\end{figure*}

The following recursive stochastic relation predicts the electron temperature "seen" by $(n+1)$th probing pulse:

\begin{equation}
\label{eq:3}
T_e(n+1)=T_0+S(n)\,\Delta T_e + \Delta T_{ph}(n+1)
\end{equation}

\begin{equation}
\label{eq:2}
\begin{aligned}
\Delta T_{ph}(n+1)=[\Delta T_{ph}(n)+S(n)\delta T_{ph}]\,exp\left(-\frac{\Delta t}{\tau_{ph}}\right)
\end{aligned}
\end{equation}

The postulated Eq.\,(\ref{eq:2}) for $\Delta T_{ph}(n+1)$ involves the modification of the "effusion process" introduced in Sec.\,\ref{tworelaxations} [Eq.\,(\ref{eq:1})] by assuming random, but constant increment of the phonon temperature at well-defined time intervals. 

We are in position to simulate the ensemble switching. We start with $P(T_e)$ dependence interpolated for 90.18$\,\mu$A. Iterating the model for $N=40\,000$ pulses does not lead to $P=1$ as we progress with simulation [Fig.\,\ref{fig:Simulation}]. The situation is different for $I_A=90.2\,\mu$A. Here, $P$ goes to 1 after sequence of heating pulses for $n$ around $32\,000$, indicating the appearance of the ensemble transition. In both cases, switching probability for each next pulse takes on a random value governed by corresponding pattern of stochastic thermal heating. The two branches of the trace in each plot indicate stochastic influence of the electron overheating: in fact the trace jumps between these two branches as we progress with simulation. Once the ensemble transition happens the electron temperature approaches its asymptotic value $T_e=596\,$mK. It tells that electrons are overheated by 96$\,$mK when the next probing pulse arrives, i.e. 2$\,\mu$s after last switching event. One should remember that electron temperature keeps changing all the time during the experiment: it goes above $T_c=1.3\,$K when the bridge it transferred to the normal state, then subsequently it relaxes towards the phonon temperature. In the discussed case the phonon temperature is 596$\,$mK after ensemble switching happened.

Our data remain in quantitative agreement with switching number transition observed for curve $\Delta \tau=2\,\mu$s, Fig.\,\ref{fig:Hysteresis}, marking the proper $n_c/N$ ratio at the onset of the ensemble switching [0.0258 in experiment - Fig.\,\ref{fig:Hysteresis}, approximately 0.025 in simulation - Fig.\,\ref{fig:Simulation}]. The repetition of the simulation with the same parameters reveals stochastic moment of the ensemble switching, as expected. For $I_A=90.20\,\mu$A switching for $N=40\,000$ is unlikely. For $I_A=90.22\,\mu$A becomes almost certain. For $I_A=90.18\,\mu$A it does not happen even for $N=100\,000$, for $I_A=90.24\,\mu$A it always appears. The range of currents, over which ensemble switching happens, agrees well with the experimental trace of Fig.\,\ref{fig:Metastable}(b), where the probing current amplitude is increased by 15$\,$nA, leading to abrupt transition. One can think about generalized \textsf{S} curve, describing the probability of the ensemble transition as a function of the probing current amplitude. Apparently, the width of such generalized \textsf{S} curve [approximately 15$\,$nA] is much narrower than the width of a usual \textsf{S} curve describing probability of switching of the bridge to the normal state [approximately 500$\,$nA, see Fig.\,\ref{fig:Calib}]. This observation supports our claim of high sensitivity of the discovered process on the control parameter, i.e. electric current, temperature or magnetic field.

The model also properly accounts for steepening of \textsf{S} curves for moderately-correlated pulses [Fig.\,\ref{fig:Scurves}(b)] when $\Delta \tau$ is long enough to allow electrons to thermalize with local phonons [$\Delta \tau > 10\,\mu$s, see the point A in Fig.\,\ref{fig:Relaxation}(b)]. In such a case consecutive switchings lead to elevated phonon temperature, and higher switching probability. The larger independent switching probability implicates more switching events and larger overheating of phonons. It is observed as the increase of the slope in the measured switching number dependencies.

\section{New perspectives for detection}
We have presented abrupt transition in the switching number $n/N$ at a critical value of the probing current amplitude and have coined it the ensemble switching. We also observed such switching upon increasing bath temperature by a fraction of miliKelvin. Similarly one may expect to see the transition for small change in magnetic flux if instead of the single junction one uses a SQUID. The abrupt transition observed in the correlated measurements means that extremely small change in the control parameter (i.e. probing current, temperature, magnetic flux) changes stochastic process dramatically: from a setting when there are rare random bunched switchings observed, the system is transformed to fully deterministic configuration when each probing pulse drives the superconductor $-$ normal metal transition. Owning to inherent thermal feedback imposed by correlations, the transition exhibits a latching effect as revealed on the hysteretic probing current dependencies.

One may argue that above features i.e. the hysteresis with discontinuity at a transition point, are desired for a detector, e.g. it can be employed for detecting magnetization reversals that produce tiny changes in magnetic flux\cite{Wernsdorfer2009}. Magnetic flux dependence of the switching probability $P(\Phi)$ in the independent regime is an \textsf{S} curve with a finite slope $dP/d\Phi$. In the correlated regime this slope becomes much steeper, thus making detector very sensitive. We speculate that such a detector would be less sensitive to electromagnetic and thermal noise since the ensemble switching involves many single switching events and appears only in response to a permanent change in the control parameter: the slow response of the phonon temperature imposes a low pass filtering on the ensemble transition. Unless the switching events are fully correlated, a single accidental switching event is not capable of driving the ensemble transition. It is in contrast to the transition edge sensors (TES) for which obtaining a stable operation in the narrow superconducting transition may be challenging\cite{Aaron2003,Horansky2013,Niwa2017}. The fast dynamical response of the TES may appear problematic in a noisy environment. The presented ensemble switching could be a detection scheme of choice when extraction of the permanent change in the monitored signal buried in the ambient noise is required. The another difference compared to the TES is opportunity to observe ensemble transition at wide range of temperatures i.e. it is possible to adjust the hysteresis for each temperature by tuning the duration between testing pulses. Having demonstrated experimentally the feasibility of the proposed scheme [Fig.\,\ref{fig:Metastable}], it remains up to future studies to verify its sensitivity, but from the analysis presented in Sec.\,\ref{fullmodel} it is clear that the width of generalized \textsf{S} curve describing probability of the ensemble transition is 1-2 orders of magnitude narrower than the width of familiar S curve describing probability of a single switching event.

\section{Discussion}
The artificial stochastic process we have created remains in strong analogy to the process which is responsible for switching superconducting wires due to multiple phase slip mechanism\cite{Bezryadin2009}. In the former case switching in a single pulse increases probability for switching during next pulses, in the latter case a single phase slip occurrence increases probability for appearing next phase slips. In both cases accumulated heating leads eventually to phase transition (we extend the meaning of phase transition to artificial process which we describe). Multi-phase slip escape is conventionally analyzed by means of mean first passage time approach, which evaluates the lifetime of the metastable state\cite{Pekker2009}. The approach allows to consider continuous stochastic process with phase slips appearing at any time. It is in contrast to our artificial stochastic process which is discrete with single switching events appearing at well-defined moments, bounded by duration of the testing pulse.

Our studies show that phonons in the current-probed nanostructures tend to be overheated with respect to the bath temperature. In our experiment we use the electric current at the level of 10-100$\,\mu$A. It is much larger value than typically used in testing tunnel junctions. However, at temperatures $\leq\,$100$\,$mK, when heat capacities are vanishingly small, even very small currents may produce significant overheating of phonons, especially in experiments conducted in steady states. This possibility is often excluded in many works by assumption of low Kapitza resistance. Even if it is correct, the assumption does not assert the thermalization of the substrate at the bath temperature.

\section{Conclusion}
Our work gives understanding of unconventional behavior of nanoscale superconducting bridges arising from the thermal feedback in the strongly correlated switching measurements. We engineer discrete stochastic process with controllable strength of correlation which may facilitate understanding of similar processes observed indirectly in nature, e.g. for the multi-phase slip driven superconductor-normal metal transition. We develop numerical recursive model incorporating stochastic heating and deterministic cooling. The model involves overheating of electrons, giving rise to nearest-neighbor correlation in the switching measurements, and overheating of phonons, accounting for long-range correlations between pulses. Our artificially produced stochastic trajectories, like the one presented in the Fig.\,\ref{fig:Metastable}, provide an interesting experimental framework for studying lifetimes of metastable states. The introduced switching protocol, using continuous pulse trains, can be used as a basis for a hysteretic detectors of magnetic flux, current and temperature when a vanishingly small, but permanent changes of these parameters are traced.

\section*{Acknowledgments}
The work is supported by Foundation for Polish Science (First TEAM/2016-1/10).

\end{document}